# Eavesdropping on Electronic Guidebooks: Observing Learning Resources in Shared Listening Environments

**Allison Woodruff, Paul M. Aoki, Rebecca E. Grinter, Amy Hurst,
Margaret H. Szymanski, and James D. Thornton**
Palo Alto Research Center
3333 Coyote Hill Road
Palo Alto, CA 94304-1314  USA

**ABSTRACT**
We describe an electronic guidebook, *Sotto Voce*, that enables visitors to share audio information by eavesdropping on each other's guidebook activity.  We have conducted three studies of visitors using electronic guidebooks in a historic house: one study with open air audio played through speakers and two studies with eavesdropped audio.  An analysis of visitor interaction in these studies suggests that eavesdropped audio provides more social and interactive learning resources than open air audio played through speakers.

**INTRODUCTION**

Previous research suggests that users of electronic guidebooks prefer *open air audio* delivered through speakers to audio delivered through a headset (see, e.g., (Kirk, 2001; Woodruff, Aoki, Hurst, & Szymanski, 2001). The well-known visitor desire for social interaction (Hood, 1983) is a key reason for this preference: when visitors use open air audio, they can listen to content together and discuss it, whereas headsets often isolate visitors into experiential "bubbles" (Martin, 2000).  However, open air audio is problematic when many visitors are present in the same location, as has been confirmed by informal experiments conducted by commercial audio guide vendors (L. Mann, Antenna Audio, personal communication).

We describe an alternative mechanism for sharing audio. This mechanism, which we call *eavesdropping*, preserves the social interaction enabled by open air audio while avoiding the audio "clutter" that open air audio necessarily entails.  In our system, visitors independently select objects in their guidebooks and listen to the audio content through one-ear headsets; these headsets allow them to hear each other speak and interact conversationally.  Further, wireless networking enables visitors to optionally listen to their companion's guidebook in addition to their own.  The intimate, often directed, nature of the resulting shared audio context has led us to call the system *Sotto Voce*.

Our design is guided by the following principle: we want to support visitor interaction with three main entities that make demands on their attention.  These entities are the information source, the visitor's companions, and the physical environment – "the guidebook, the friend, and the room" (Woodruff, Aoki et al., 2001).  As we add capabilities that enhances visitor interaction with one entity, we must be careful that we do not compromise visitor interaction with other entities (e.g., we do not want to improve visitor-visitor interaction at the expense of visitor-room interaction.)

To understand the impact of the eavesdropping mechanism on the overall visitor experience, we conducted two studies of visitors using the system to tour a historic house.  We applied qualitative methods to the resulting data, including an analysis of visitor interviews and an applied conversation analytic study of recorded audiovisual observations.  Because the eavesdropping was an optional feature that visitors could turn on or off at will, we observed several categories of use, e.g., pairs of visitors who did not use eavesdropping, pairs of visitors who used eavesdropping intermittently, and pairs who engaged in continuous *mutual eavesdropping*.

In this paper, we focus on the visitors who engaged in mutual eavesdropping, which is the category that most closely approximates open air audio.  We compare the typical behavior of these mutual eavesdroppers to that of visitors in a previous study who used open air audio to create a shared listening experience (Woodruff, Aoki et al., 2001; Woodruff, Szymanski, Aoki, & Hurst, 2001). (The three studies are summarized in Table 1.)  Most of the discussion is based on analysis of the observational data. We observe that mutual eavesdroppers had a different activity structure and were more mobile than visitors who used open air audio.  As a result of these changes, mutual eavesdroppers had increased resources for engaging in interactive learning: they had richer and more extensive social interaction, and they had more resources for physically exploring their environment.  For example, visitors had more substantive discussion in response to

| Table 1. Summary of studies conducted. |||||
|---|---|---|---|---|
| Study | Audio sharing mechanism | Participants ||  Previous papers |
| | | Recruited | Public | |
| 1 | Open air | 14 | | (Woodruff, Aoki et al., 2001; Woodruff, Szymanski et al., 2001) |
| 2 | Eavesdropping | 12 | | (Aoki et al., 2002) |
| 3 | | | 47 | -- |

guidebook descriptions, and they were more likely to discuss objects not described in the guidebook. Given the importance of social learning in the museum environment (Falk & Dierking, 2000), the preliminary evidence presented here is encouraging and suggests further avenues for work along these lines.

The remainder of the paper is organized as follows. First, we discuss the design of *Sotto Voce*. Next, we describe the method employed in our user study. We then turn to findings. These are divided into the impact of the design on visitor behavior and the implications of these behavioral changes for visitors' learning resources. After discussing related work, we summarize our findings and describe future directions.

**PROTOTYPE DESIGN**

In this section, we discuss the design and implementation of the guidebook device, key aspects of its user interface, the design goals for the audio environment, the eavesdropping mechanism, the audio delivery mechanism, and the construction of the audio content. The design is the same as that used in Study 2, reported in (Aoki et al., 2002), but we briefly discuss it here to provide context. Overall, visitors have a positive response to the guidebook and report that it is easy to use (Aoki et al., 2002; Woodruff, Aoki et al., 2001).

*Guidebook device*. We implemented the device using the Compaq iPAQÔ 3650 handheld computer, which includes a color LCD touchscreen display. With an IEEE 802.11b wireless local-area network (WLAN) card, the device measures 163mm x 83mm x 34mm (6.4" x 3.3" x 1.3") and weighs 368g (13 oz.).

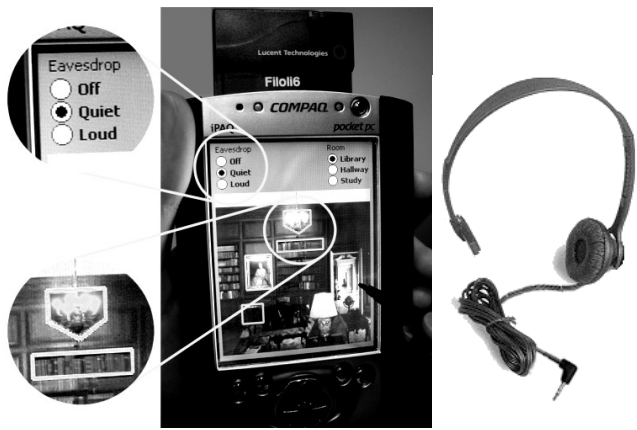

**Figure 1. Electronic guidebook and headset.**

To support eavesdropping, paired devices communicate over the WLAN using Internet protocols (UDP/IP). The audio content is the same on all devices, so the devices send and receive control messages ("start playing clip 10," "stop playing clip 8") rather than waveform audio. Since our goal is to enhance co-present interaction, the device does not support remote voice communication.

*User interface*. This part of the system is very similar to that used in previous studies, and its design rationale is more thoroughly described elsewhere (Woodruff, Aoki et al., 2001). Individual visitors obtain information about objects in their environment using a visual interface. This helps visitors maintain the flow of their visual task (looking at the room and its contents), which tends to reduce demands on user attention. The interface resembles a set of Web browser imagemaps; at a given time, the visitor sees a single photographic imagemap that depicts one wall of a room in the historic house (Figure 1, center). Visitors change the viewing perspective (i.e., display a different imagemap) by pressing a hardware button. When visitors tap on an imagemap target, the guidebook plays an audio clip that describes that object. Many, but not all, of the objects visible on the screen are targets; to help visitors identify targets, the guidebook displays *tap tips* (Aoki, Hurst, & Woodruff, 2001) – transient target outlines that appear when the user taps and fails to "hit" a target (Figure 1, bottom left). A demonstration of the visual interface is available online (http://www.parc.com/guidebooks/).

*Audio design goals*. Results from Study 1 suggested several design criteria. Visitors want to be able to share audio descriptions and converse. At the same time, visitors want to retain personal control over the selection of descriptions. Further, the design needs to facilitate the ability of visitors to explore their physical environment, and the design needs to be sufficiently lightweight that it makes minimal demands on the users' attention. Finally, the design needs to be feasible in public environments with many visitors. These criteria ruled out a number of options like open air audio (which is not feasible for large numbers of visitors) or splitters that allow two visitors to listen to audio from a single device (which restrict visitor movement and do not allow visitors individual control over the audio content to which they are listening). The eavesdropping model described below is an alternative that meets all of the criteria.

*Eavesdropping*. In concrete terms, paired visitors share audio content as follows. When visitor A selects an object on her device, she always hears her own audio clip. If A is not currently playing an audio clip, but her companion B is, then B's audio clip can be heard on A's device. In other words, audio clips are never mixed, and A's device always plays a personal clip (selected by A) in preference to an eavesdropped clip (selected by B). Audio playback on the paired devices is synchronized; if A and B are both listening to their own clips and A's clip ends first, A will then hear the remainder of B's clip as if it had "started in the middle." To control a device's eavesdropping volume (i.e., the volume at which A hears B's clips), the interface includes three option buttons: "Off," "Quiet" and "Loud" (Figure 1, top left). "Loud" is the same as the volume for personal clips.

In abstract terms, eavesdropping provides a relatively simple *audio space* model (Mackay, 1999). We did consider other options, such as a telephony-like connection model in which visitors would independently initiate and terminate audio sharing sessions with their companions. We also considered email-like asynchronous models in which visitors would send and receive audio clips at their convenience. We rejected more complex abstractions that involved multiple actions (send/receive, connect/accept/reject, etc.) because we believed that the necessary interface gestures would distract visitors from their experience with the environment and their companions. In the audio space model, sharing requires no gestures of its own. To "receive," a visitor merely sets the eavesdropping volume. To "send," a visitor simply selects an object; playing a description has the side effect of sharing it, if the companion chooses to eavesdrop. The audio space model has the further advantage that it supports simultaneous listening, which enhances social interaction by creating the feeling that the content is part of a shared conversation (Woodruff, Szymanski et al., 2001).

*Audio delivery*. Visitors hear descriptions through headsets. We conducted a small study (n=8) to identify headsets that would allow visitors to converse and that visitors would readily accept (Grinter & Woodruff, 2002). Based on this study, we chose commercial single-ear telephone headsets, locally modified by the removal of the boom microphone (Figure 1, right). This configuration leaves one ear available to hear sounds from the external environment, and visitors find the over-the-head design desirable because it is familiar and gives them the sense that the headset is securely attached.

*Audio content*. The prototype contains descriptions of 51 objects in three rooms of the house. In most regards, the descriptions are recorded along principles described in (Woodruff, Szymanski et al., 2001). The audio clips vary in length between 5.5 and 27 seconds, with the exception of one story that runs for 59 seconds. The clip length is much shorter than conventional audio tour clips, which often run to 180 seconds, and is intended to facilitate conversation by providing frequent opportunities for visitors to take a conversational turn.

Since we use single-ear headsets, both personal and eavesdropped audio content are necessarily presented in the same ear. We distinguish the two types of content using two mechanisms. First, we apply a small amount of reverberation to the eavesdropped audio. A single earphone cannot effectively deliver spatialized audio (Blauert, 1997), but can support other sound effects; we chose reverberation after conducting user tests (n=6) involving scenario-based tasks using the guidebook. Second, the default eavesdropping volume ("Quiet"), which is most frequently used by visitors, is softer than the personal volume.

**METHOD**

We have conducted three major user studies at Filoli, a Georgian Revival historic house located in Woodside, California (http://www.filoli.org/). Study 1 used an earlier version of *Sotto Voce* that supported open air audio, whereas Studies 2/3 used the current version of *Sotto Voce* that supports eavesdropping as described in the design section of this paper. Study 1 and Study 2 involved previously recruited participants on days the house was closed to the general public, whereas Study 3 involved 47 visitors recruited on-site on days the house was open to the general public. (Again, these studies are summarized in Table 1.)

Because the participants and procedures for Study 1 and Study 2 have been reported previously, below we report only the participants and procedure for Study 3. We then discuss our analytic methods, which were the same in all studies.

*Participants*. In Study 3, we observed 20 pairs, one group of three, and one group of four using the guidebooks. These pairs and larger groups were comprised of visitors who had come to Filoli together, e.g., mother/daughter or friend/friend pairs. The majority of visitors had not previously used a handheld device. The visitors covered a wide range of ages: the youngest visitors were in the "18-29" age range, and seven visitors who used the guidebook were "over 70." (While we had several children test *Sotto Voce* in the first and second studies, visitors from the ages of approximately 5-17 are quite rare at Filoli unless they are visiting with a school group.)

*Procedure*. Visitors to the house were recruited at the entrance to the Library, the first room discussed in the guidebook. After signing consent forms, visitors were fitted with a wireless microphone, given guidebooks, and trained in their use. Next, they visited the three rooms for which the guidebook had content. When they finished using the guidebooks, they participated in a semi-structured interview.

The visitors' conversation and comments during the interview were recorded using the wireless microphones,

the visitors were videotaped by fixed cameras while using the guidebooks (all visitors to the house were notified that videotaping was in progress), and the visitors' use of the guidebooks was logged by the device.

Visitors typically spent about 15 minutes using the electronic guidebooks. Their participation in the study took approximately 30-45 minutes; no time limits were imposed during any portion of the procedure.

*Analysis.* We analyzed the data in several ways. For example, we transcribed and analyzed the interview data to examine the visitors' attitudes and feelings about the technology and their experience. The majority of the findings presented in this paper are based on another method we used, *conversation analysis* (Sacks, 1984).

Conversation analysis is a sociological method used to examine naturally occurring social interaction to reveal organized patterns. To find such patterns, conversation analysts study collections of interactive encounters and identify *sequences* of actions that were recurrently made by the participants. Actions in our context might include making a verbal utterance, pointing at an object, or selecting a description.

To this end, we create a composite video of visitors and their guidebook screens and audio (re-created from the guidebook activity logs). We then transcribe the actions taken by visitors, including dialogue, and look for recurring patterns to identify visitors' systematic practices.

**FROM OPEN AIR TO EAVESDROPPING: CHANGES IN VISITOR BEHAVIOR**

In this section, we compare the behavior of the pairs who chose to use mutual eavesdropping in Studies 2/3 to that of similarly engaged pairs who used open air audio in Study 1.

Specifically, we discuss the structure of the visitors' interactions and their physical mobility. The effect of these aspects can be identified in the visitors' learning-related behavior, which is the subject of the following section.

**Changed Activity Structure**

Visitor activity was structured very differently with eavesdropped audio than with open air audio. The new structure had a lower coordination cost, demanding less attention. The decreased attention burden was reflected in the visitors' interactions.

In all of the studies, a single overall structure pervaded the interactions. Specifically, they exhibited the sequential, multi-phase organization known as *storytelling* in the conversation analytic literature (Sacks, 1974); as part of this organization, visitors created a conversational role for the audio descriptions, i.e., they treated the guidebook like a "third party" taking an extended conversational turn (Aoki et al., 2002; Woodruff, Szymanski et al., 2001). Paired visitors entered a state of *engagement* at the beginning of a given storytelling sequence; levels of engagement generally rose and then fell over the course of a given sequence; and visitors then had the options of dis-engaging (resulting in independent activity), remaining engaged in shared activity, or maintaining a nascent engagement in expectation of subsequent re-engagement (Szymanski, 1999).

With open air audio, visitor interactions tended to focus on choosing individual objects and coordinating with their companions to listen to the descriptions. This setup, repeated for each sequence, focused more attention on coordination activity than seems necessary or desirable. However, the open air audio did afford the opportunity to participate in shared responses to the "story," motivating the visitors to begin setup for another sequence.

By contrast, participation in mutual eavesdropping created an ongoing assumption that the couple would continue in the shared activity. This supposition of continuing shared activity meant that setup tended to be cursory. Further, while open air audio was primarily conducive to follow-up discussions that related directly to descriptions, mutually eavesdropped audio was conducive to many diverse types of follow-up sequences such as discussion of objects not described in the guidebook.

The change in activity structure had at least two beneficial effects. First, by reducing the effort needed to choose and listen to descriptions, mutual eavesdropping freed visitors to direct more attention to meaningful interactions with their environment and their companions (i.e., away from the guidebook and routine coordination). In other words, the reduction in low-quality coordination talk meant that visitors had more time to investigate the room and its contents and that a higher proportion of talk tended to focus on topics of substance. Second, since the new activity structure supported more diverse types of sequences, visitors were more likely to pursue new topics or investigate objects not described in the guidebook.

**Increased Mobility**

Visitors in Studies 2/3 were noticeably more mobile during periods of engagement. In Study 1, the open air audio was played at a low volume, so any movement that changed the relative position of the visitors could cause significant sound attenuation due to distance or blockage (e.g., due to interposed obstacles – even changes in body orientation could cause the audio to be blocked). As a result, couples tended to remain close together and stationary while sharing audio descriptions. See Figure 2a, in which a grandmother is bending over to listen to the audio description that her granddaughter is playing of the portrait over the fireplace. Note how this position prevents her from examining the painting while she listens. In Studies 2/3, visitors were less constrained. Because movement could not attenuate the audio information, visitors could separate from each other physically while listening to descriptions and remaining engaged. See Figure 2b, in which both visitors are listening to a description of the marble staircase. While both visitors are examining the staircase, they have each chosen different vantage points. However, this positioning does not compromise their social connection: when the audio description reveals that only the first four steps are actually solid marble, the male visitor looks to his companion and she laughs, even though they are not standing together.

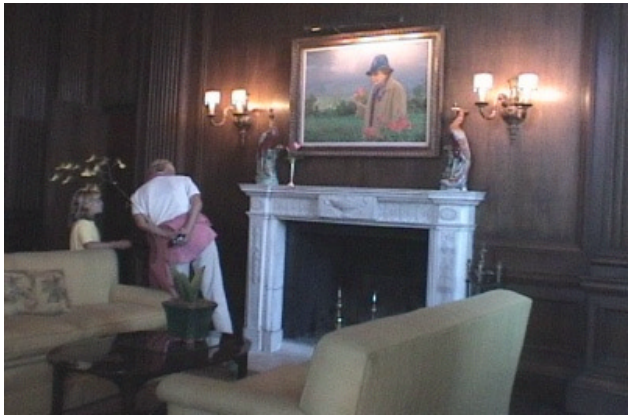 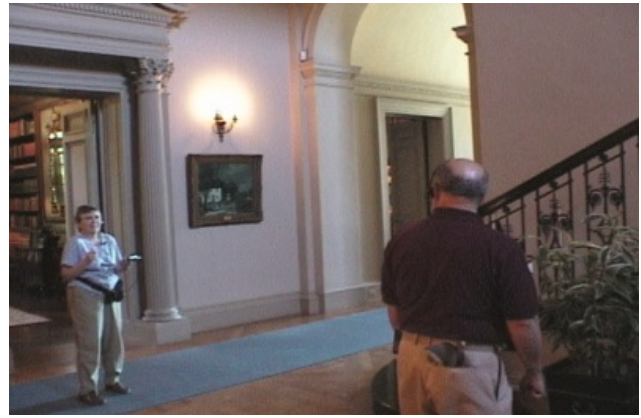

| (a) Visitors standing close together when using open air audio. | (b) Visitors standing far apart when using mutually eavesdropped audio. |

**Figure 2. Comparison of visitor mobility patterns.**

The increased mobility resulting from use of mutual eavesdropping took many forms. We observed several common behaviors that rarely, if ever, occurred with open air audio. For example, visitors would often walk together while a description was playing, e.g., to approach the object being described. In other cases, a single visitor would walk closer to the object currently being described while their companion remained stationary. In still other cases, a visitor would investigate a different object from the one currently being described and then rejoin their companion.

## FROM OPEN AIR TO EAVESDROPPING: INCREASED RESOURCES FOR LEARNING

The study observations provide evidence that both of the factors described in the previous section – the changed activity structure and increased mobility during engagement – improved the learning environment. Here, we discuss two learning-related resources that were enhanced by the guidebook: the nature of the visitors' social interaction and their opportunities for exploring the room and its contents.

Our analyses are based upon a collection of transcribed excerpts, from which the following extracts have been derived. These extracts are meant to exemplify and highlight specific behaviors rather than to illustrate the organization of the visitors' interactions (space limitations preclude the use of representative excerpts of this kind).

Table 2 summarizes the notation used in this section. For clarity of presentation, the extracts have been simplified to use conventional capitalization and punctuation (e.g., commas and periods).

Note that the discussion in this section is limited to the *increased availability* of *learning resources*. A claim of *increased learning* would require a different type of study, e.g., one that measured the visitors' knowledge before and after their visit. Such a study is beyond the scope of this paper.

| Table 2. Summary of transcription notation. | |
|---|---|
| **X:** <br> **X-PDA:** | Visitor X is speaking. **((**comment or action**))** <br> *Visitor X's guidebook is speaking.* |
| °°whisper°° | Speech at reduced volume. |
| <u>em</u>phasis | Emphasis in speech. |
| **(***n***)** | A conversational pause of *n* seconds. **(.)** indicates a "micropause." |

### Depth and Length of Social Interaction

When using mutual eavesdropping, visitors responded more fully to audio descriptions. Visitors were also more likely to discuss features of the object not mentioned in the description or to discuss objects that were not described in the guidebook at all.

Each of these phenomena represents a way in which visitors collaboratively built on the shared audio descriptions, working together to construct mutual learning resources that broaden, deepen or expand their discussion of the room's contents. The importance of such social learning, particularly (but not limited to) conversation, has been widely supported in the visitor studies literature (see, e.g., (Falk & Dierking, 2000; Russell, 1994)). Social interaction around artifacts affords the "opportunity for the visitor to make connections with familiar concepts and objects" (Hein, 1995); adding resources for interaction adds more such opportunities. The remainder of this subsection gives some examples, linking the behavioral changes of the previous section to the construction of learning resources.

*Characterizations*. With mutual eavesdropping, response to an audio description was likely to be more reflective and include a physical focus relative to the object being described. With open air audio, visitors would often have a very minimal response, and even the more substantive responses were generally limited to reactions to the description that had just occurred.

The following extracts are representative of this effect. Consider V and W (Extract I, Study 1) who were listening with open air audio.

| Extract I. | |
|---|---|
| V-PDA: | *Many of the top shelves contain false books. They are lighter than normal books, so they reduce the stress on the bookcases. Many are made of greeting cards, clothing, fabric, et cetera.* |
| W: | Eh hah, that's a riot. ((W looks at V and smiles)) (0.2) They're just for looks. |

While V and W do share a response, the substance is limited to a single paraphrase of the audio description, analogous to "text echo" of exhibit labels (McManus, 1989) (though possibly more affective, because of the audio delivery).

By contrast, consider an interaction in which J and L (Extract II, Study 2) were mutually eavesdropping.

| Extract II. | |
|---|---|
| L-PDA: | *… All of the architectural features of this room, including the walnut panelling, are modelled on an 18th century British library. In the original library, each of the outlined panels would have contained framed pictures.* |
| L: | Re<u>al</u>ly. |
| J: | Yeah. |
| L: | That's a lot of pictures. ((points at wall and sweeps arm across walls)) |
| J: | <u>That</u>'s a lot of pictures. ((nods "yes")) |
| | (.) |
| J: | °°That would've been very cluttered.°° |

This interaction is richer in many respects than that shown in the previous extract. J and L both speak, taking more turns to discuss the description than V and W. By making a statement about the number of pictures, L reinforces for J a quantitative observation that is not made in the description itself. By gesturing at the many empty wall panels, L adds physical, spatial, and visual elements to the experience, linking both visitors to a vision of the "original" library that overlays their actual surroundings. J agrees with the quantitative statement, pauses, and then responds by saying it would have been "cluttered," a qualitative assessment that indicates that she has in fact visualized the room as it might have been.

This increased reflection on descriptions was evidenced in many ways. Visitors worked together more to understand descriptions, e.g., a visitor would sometimes express confusion about a description and their companion would help them understand it. Further, with mutual eavesdropping, visitors more frequently branched off into sequences that were not directly related to the description of the content. For example, they might point out a specific physical feature of the object that was not mentioned in the description, or discuss some way that it related to their own life.

Additionally, visitors showed more evidence of establishing complex relationships between objects. One visitor pointed at a series of paintings on different walls, saying, "Okay, so that's his wife, and that's his mother, right?" Or consider T's comments (Extract III, from Study 2) when he first enters a particular room. His statements indicate that he has constructed a category of "secret cabinets" that occur in this house and that he is alert to instances of this category as he moves from room to room.

| Extract III. | |
|---|---|
| T: | Ah, more secret cabinets. |
| | (0.4) |
| T: | I like that <u>a</u> <u>lot</u> about this house. ((walks into the bar closet)) |

Interactions displaying this kind of orientation – i.e., at the granularity of a thematic collection rather than a single object – almost never occurred with open air audio.

Moreover, mutually eavesdropping visitors often discussed objects that were not described in the guidebook, unlike open air audio visitors. The following sequence, in which J teaches L about a plant, occurred immediately after they finished their response to a description:

| Extract IV. | |
|---|---|
| J: | Okay, your- your test for the day, what's that one? ((points to plant)) |
| | (0.2) |
| J: | The plant. |
| | (0.4) ((L leans in to look)) |
| L: | Morning glory. Eh heh heh heh, I don't know, what is it? |
| J: | I think it's a mandevilla vine, but I'm not sure. |
| L: | °°God, I can't believe you know that.°° |

*Reasons.* Both of the behavioral changes resulting from use of mutually eavesdropped audio had impact on social interaction. The primary factor was the new activity structure, which allowed more space for reflection and for visitors to initiate new conversational sequences that were not structured around the audio descriptions. Increased mobility constituted a secondary factor. Visitors would often start descriptions while they were far away from objects. As mentioned above, visitors were unlikely to walk toward the object while the description was playing with open air audio. However, with eavesdropped audio, they were more likely to approach the object; being close to the object when the description ended gave them more opportunities to observe and discuss its specific features.

**Expanded Resources for Physical Exploration**

With mutually eavesdropped audio, the examination of objects was more frequently occasioned by their presence in the *room* rather than their presence in the *guidebook*. Once visitors began to examine an object, they might discuss it or play a description of it if one were available.

This implicit shift in emphasis from the guidebook to the room as the impetus for exploration is important because it shifts the visitor's role. It is broadly (though perhaps not universally) accepted that learning is enhanced by enabling visitors to navigate the museum without leading them through it (Falk & Dierking, 2000). However, even "free choice" navigation can be constrained by, e.g., which objects have descriptive content associated with them. Visitor behavior indicates that use of mutual eavesdropping increased the guidebook's utility as a reference (an adjunct to the room) as opposed to an inventory (a directed guide to the room).

*Characterizations.* In the study using open air audio, examination of objects often began with objects contained in the guidebook and proceeded by spatial locality. That is, visitors tended to switch the visual interface to a given wall and then look at the objects in the guidebook that interested them on that wall. Object choice was often based on targets seen in the visual interface or on short-term memory of such targets.

In the eavesdropping studies, the next object to examine was less frequently chosen based on availability in the guidebook. (In many of these cases, we know that the examination was prompted by the room rather than the guidebook because the objects were not described in the guidebook. In the other cases, the visitors spoke their thoughts aloud – which was entirely self-prompted since none of the studies involved a speak-aloud protocol.) Instead, visitors would encounter objects in their field of view, e.g., objects that were near an object they had just examined, or they would deliberately examine sequences of objects they perceived as being related.

For example, in Extract III, T walks into a new room, notices the bar closet and actually walks into it. *After* this, his companion D finds the description in the guidebook and plays it. Note that because the sound does not attenuate, the visitors can listen to the description together while T stands inside the tiny closet and D stands outside.

*Reasons.* While the same resources were available with open air audio, they were used much more frequently in the mutual eavesdropped case due to the changed activity structure and the increased mobility in the room. Specifically, the mutually eavesdropped audio was more conducive to sequences that were not directly responsive to guidebook content; visitors were generally more open to external triggers with the new activity structure. Visitors acted in a manner more consistent with "Let's see what's here in the room" than with "Let's see what's here in the guidebook." Further, visitors had more attention to give to the room due to the reduced attentional demands and wandered more in the room due to increased mobility, so they were more likely to encounter and investigate objects.

**RELATED WORK**

Our work draws together three main areas of research. Space limitations preclude an extended discussion; additional references are contained in (Aoki et al., 2002; Woodruff, Aoki et al., 2001; Woodruff, Szymanski et al., 2001).

*Interaction in museum settings.* The importance of social interaction to museum visitors is well known (e.g., (Hood, 1983)). There are two types of studies of particular interest. McManus observed visitor usage of text labels; she noted that visitors were inclined to treat exhibit labels *as conversation* to which they had been party (McManus, 1989). A number of studies of museum visitors have been conducted using methods derived from conversation analysis (see, e.g., (Falk & Dierking, 2000), Ch. 6, and (vom Lehn, Heath, & Hindmarsh, 2001). These studies focus on talk, interaction and learning in conventional environments; here, we have focused on the effects of electronic guidebooks on social interaction and learning resources.

*Electronic guidebooks.* The cultural heritage community has formally studied electronic guidebooks for many years (Screven, 1975). Related work in HCI has focused on aspects such as, e.g., location-aware computing (Abowd et al., 1997) and only recently have significant user studies been reported (e.g., (Cheverst, Davies, Mitchell, Friday, & Efstratiou, 2000)). The HCI studies focus on system design and evaluation; here, we focus on the effects of our system on visitor interaction.

*Media and interaction.* There is an extremely rich literature on collaborative multimedia environments; of particular interest are media spaces (Mackay, 1999). Many of these systems have been evaluated, but most apply either ethnographic techniques or quantitative methods to studies of installed workplace systems. In this study, we apply conversation analytic techniques to the study of a mobile, leisure-activity system that provides shared access to application content.

**CONCLUSIONS**

In this paper, we have described an eavesdropping mechanism that allows visitors to listen to each other's guidebooks. Our findings show that mutual use of this eavesdropping mechanism can lead to increased learning resources as compared with the use of speakers in open air: couples using mutual eavesdropping in Studies 2/3 had more substantive interactions and exhibited an increased awareness of the room and its contents when compared to those using open air audio in Study 1.

New work is addressing some of the open issues from this study. We are preparing a discussion of the ways in which the visitors creatively used our eavesdropping mechanism for tasks other than enhancing their social interaction, e.g., for monitoring their children. We are also planning an experiment using bone conduction headsets that can provide binaural audio without occluding the ears.

**ACKNOWLEDGMENTS**

We are deeply grateful to Tom Rogers and Anne Taylor of Filoli Center for their assistance with this project. Amy Hurst performed this work during an internship from the College of Computing, Georgia Institute of Technology.